\documentclass[11pt,dvips]{article}
\usepackage{epsfig,times} 
\usepackage{picinpar}
\usepackage{wrapfig}
\usepackage{floatflt}
\usepackage{spez}
\makeatletter
\renewcommand{\section}{\@startsection{section}{1}{0in}
	{0.4\baselineskip}{0.1\baselineskip}{\Large\bf}}
\renewcommand{\subsection}{\@startsection{subsection}{2}{0in}
	{0.25\baselineskip}{-\baselineskip}{\large\bf}}
\renewcommand{\subsubsection}{\@startsection{subsubsection}{3}{0in}
	{0.1\baselineskip}{-\baselineskip}{\normalsize\bf}}
\makeatother
\newcommand{\ep}{\epsilon}
\newcommand{\per}{\scriptscriptstyle \bot}
\newcommand{\md}{{\rm d}}
\newcommand{\rgp}{r_{{\rm g} \per}}

\newcommand{\tauesc}{\tau_{\rm\scriptscriptstyle esc}}
\newcommand{\kappaB}{\kappa_{\rm\scriptscriptstyle B}}
\newcommand{\nuesc}{\nu_{\rm\scriptscriptstyle esc}}
\newcommand{\eqb}{\begin{eqnarray}}
\newcommand{\eqe}{\end{eqnarray}}

\begin{document}
\thispagestyle{myheadings}
\markright{OG 3.3.20}
\begin{center}
{\LARGE \bf Ion injection and acceleration at modified shocks}
\end{center}
\begin{center}
{\bf U.D.J. Gieseler$^{1}$, T.W. Jones$^{1}$ and Hyesung Kang$^{2}$ }\\
{\it $^{1}$University of Minnesota, Department of Astronomy,
        116 Church St. S.E., Minneapolis, MN 55455, U.S.A.\\
$^{2}$ Department of Earth Sciences, Pusan National University,
       Pusan 609-735, Korea}\\
\end{center}
\begin{center}
{\large \bf Abstract\\}
\end{center}
\vspace{-0.5ex}
The theory of diffusive particle acceleration explains the spectral 
properties of the cosmic rays below energies of $\simeq 10^{15}$ eV 
as produced at strong shocks in supernova remnants (SNR's). 
To supply the observed flux of cosmic rays, a significant 
fraction of the energy released by a supernova has to be transfered to 
cosmic rays. The key to the question of the efficiency of SNR's in producing
cosmic rays is the injection process from thermal energies. A self-consistent 
model has to take into account the interaction of the accelerated particles 
with magneto-hydrodynamic waves, which generate the particle 
diffusion, a requisite for the acceleration process. Such a nonlinear 
model of the turbulent background plasma has been developed recently 
(Malkov 1998). We use this model for the first numerical
treatment of the gas dynamics and the diffusion-convection equation at a 
quasi-parallel strong shock, which incorporates a plasma-physical 
injection model to investigate the cosmic ray production.
\vspace{1ex}

\section{Introduction}
\label{intro.sec}
The problem of the efficiency of particle acceleration at shocks by
the first order Fermi acceleration process, and the strength
of the back-reaction of these particles on the plasma flow, is intimately 
related to the injection process. This describes the rate at which particles
are not only part of the thermal plasma, which is compressed and heated
when it passes the shock, but become subject to energy gain due to the Fermi
process, described by the diffusion-convection equation (e.g.~Skilling 1975).
We will follow closely a picture of this injection process which has been 
developed by Malkov \& V\"olk (1995) and Malkov (1998).

The spatial diffusion of particles, which is an essential part of their
acceleration process,
is produced by magneto-hydrodynamic waves, which are in turn generated by 
particles streaming along the magnetic field $\vec{B}_0$. We will 
assume the field $\vec{B}_0$ to be parallel to the shock normal 
($x$-direction). The magnetic field, which corresponds
to a circularly polarised wave can be written as
$\vec{B}=B_0\vec{e}_x + B_{\per}(\vec{e}_y \cos k_0x -\vec{e}_z \sin k_0x)$.
The amplitude $B_{\per}$ will be amplified downstream by a factor
$\rho/\rho^*=r$,
where $\rho$ and $\rho^*$ are the plasma densities downstream
and upstream respectively and $r$ is the compression ratio. 
The downstream field can be described by a parameter $\ep$, for which we 
assume 
$\ep := B_0/B_{\per}\ll 1$, in the case of strong shocks considered here.
Note that the perpendicular component of the magnetic field leads effectively
to a quasi-alternating field downstream of the shock for particles
moving along the shock normal. Only particles with a gyro radius
 $\rgp=p\,c\,\sin\alpha/(eB_{\per})$ 
for which the condition $k_0 \rgp > 1$ is fulfilled, would be able to 
have a net velocity with respect to the wave frame, i.e.~the downstream 
plasma would be transparent. The fraction of these
particles, which are also in the appropriate part of the phase space 
(depending on the shock speed) would be able to cross the shock from 
downstream to upstream. For the protons
of the plasma, the resonance condition for the generation of the plasma waves
gives $k_0 v_{\rm p} \approx \omega_0 = \omega_{\per}B_0/B_{\per}$, where 
the cyclotron frequency of protons is given by
 $\omega_{\per}=eB_{\per}/(m_{\rm p} c)$, and $v_{\rm p}$ is the mean 
downstream 
thermal velocity of the protons. We now have for the thermal protons
 $k_0 \rgp \approx \ep \ll 1$, which means, that most of the downstream 
thermal
protons would be confined by the wave, and only particles with higher velocity
(the tail of the Maxwell distribution) are able to leak through the shock.
Ions with mass to charge ratio higher than protons, have a proportional larger
gyro radius, so that the injection efficiency of protons, would yield a lower 
limit for the ions.
On the other hand, for thermal electrons a plasma with
such proton generated waves would not be transparent. However, reflection
off the shock could become efficient.
In the following we will focus on the wave generating protons.

To find the part of the thermal distribution for which the magnetised plasma 
is transparent, and therefore forms the injection pool, Malkov (1998) solves
analytically the equations of motion for protons in such self generated waves.
This is a nonlinear problem, because of the feedback of the 
leaking particles on the transparency of the downstream plasma, mediated by the
waves generated in the upstream region, which are swept to downstream
with the plasma flow. He finds a transparency function $\tauesc$ which 
depends mainly on the particle velocity $v$, the velocity of the 
shock in the downstream plasma frame 
$u_2$ and the parameter $\ep$. This function expresses the
fraction of particles which are able to leak through the magnetic waves,
divided by the part of the phase space for which particles would be able to 
cross 
from downstream to upstream when no waves are present. Furthermore, as a 
result of the above described feedback, he is able to constrain the 
parameter $\ep$, leaving essentially no free parameter.

The plasma flow structure, of course, also depends on the cosmic rays. 
These provide an energy sink for the plasma and 
smooth the shock structure due to a gradual deceleration of the plasma 
flow. We use a numerical method of solving the gas dynamics equations
together with the cosmic ray transport equation which has been developed by
Kang \& Jones (1995), and which is outlined in more detail in 
Sect.~\ref{method}. We show in the next section how we incorporate the above 
described model in this numerical method.

\section{Injection model}
\label{model}
The key part of the solution to the problem of proton (ion) injection
is the above described transparency function of the plasma. Kang \& Jones 
(1995) used a numerical injection model with two essentially 
free parameters which describe boundaries in momentum at which
particles can be accelerated and from which these contribute to 
the cosmic ray pressure. The transparency function provides now a 
more physical definition of exactly this intermediate region. 
For the adiabatic wave particle interaction it is given by Malkov (1998)
Eq.~(33), with $\tauesc=2\,\nuesc/(1-u_2/v)$, where the wave escape function
$\nuesc$ is divided by the fraction of the available phase space. 
We use here the following approximation:
\eqb\label{tau}
\tauesc(v,u_2)=H\left[ \tilde{v}-(1+\ep) \right]
        \left(1-\frac{u_2}{v}\right)^{-1}\,
        \left(1-\frac{1}{\tilde{v}}\right)\,
 \exp\left\{-\left[\tilde{v}-(1+\ep)\right]^{-2}\right\}\,,
\eqe
\begin{wrapfigure}[15]{r}{8.3cm}
\vspace{5.2cm}
\includegraphics{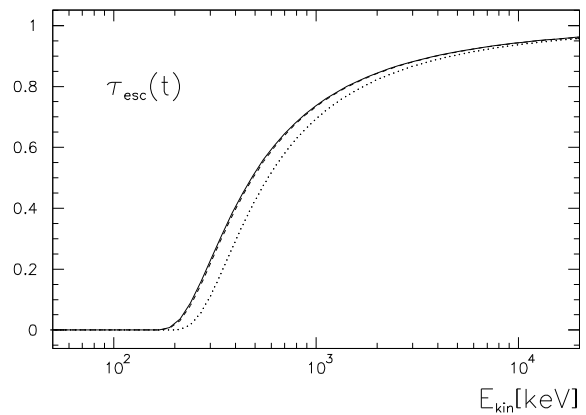}
\caption{\it Escape function Eq.~(\ref{tau}), for protons 
at times $t/t_0=0$ (dotted), 5 (dashed) and  10 (solid line).}
\label{tau_plot}
\end{wrapfigure}
where the particle velocity is normalised to $\tilde{v}= vk_0/\omega_{\per}$ 
and $H$ is the Heaviside step function.
We argued above, that $\omega_{\per}/k_0\simeq u_2/\ep$ 
(see Malkov 1998, Eq.~42).
The transparency function now solely depends on
the shock velocity $u_2$, the particle velocity
$v$, and the relative amplitude of the wave $\ep$. An important result of
Malkov (1998) is the constraint on the parameter $\ep$. He found
$0.3\la\ep\la 0.4$, as a result of the feedback described above.
Comparison with hybrid simulations suggest $0.25\la\ep\la 0.35$ 
(Malkov \& V\"olk 1998).
The function (\ref{tau}) is plotted in Fig.~\ref{tau_plot} 
for $\ep=0.35$ and protons vs.~their kinetic energy for three different 
times during the evolution of a modified shock (see below). The time dependence 
arises from the modification of the downstream plasma velocity by the cosmic
rays. We use this function to correct the result of the cosmic ray transport
equation for the upstream phase space density after each numerical time step. 
That means, the Maxwell distribution is restored (according to the
appropriate plasma temperature) where $\tauesc=0$, because here the cosmic ray
acceleration has no effect. For higher velocities we multiply the
difference between the new and old phase space distribution by the 
transparency function. Where $\tauesc=1$, the solution of the transport
equation is unchanged, because for these particles the plasma is completely
transparent, and all of them are subject to first order Fermi acceleration.
The transition between these regions is then described by the shape
of the transparency function (\ref{tau}). The phase space distribution then 
changes gradually (in energy) from a Maxwell distribution to a power law 
distribution at high energies, and it is the difference between this 
distribution and the Maxwell distribution, which we use to calculate the 
cosmic ray pressure $P_{\rm c}$.

\section{Numerical method and results}
\label{method}
In order to find the time evolution of the cosmic ray energy spectrum,
we solve the time dependent cosmic ray transport equation 
(using an implicit Crank-Nicholson scheme), together with the general 
equations of gas dynamics (using a TVD code, see Harten 1983),
including the cosmic ray pressure $P_{\rm c}$ and the energy flux $S$ which 
couples these equations. We refer to Kang \& Jones (1995) for a more detailed
description, and emphasize here only the main differences with that work.   
Very important for the injection process is the 
\label{results}
\begin{wrapfigure}[21]{r}{8cm}
\vspace{8.2cm}
\includegraphics{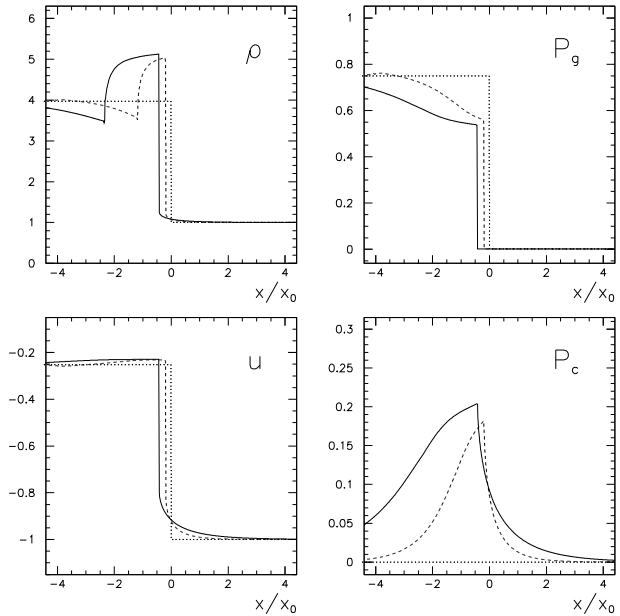}
\caption{\it Gas density $\rho$, pressure $P_{\rm g}$, velocity $u$, and cosmic
ray pressure $P_{\rm c}$, at times $t=0$ (dotted), $t=5\,t_0$ (dashed) and 
 $t=10\,t_0$ (solid line).}\label{plasma}
\end{wrapfigure}
energy transfer between plasma and cosmic rays. 
The injection energy-density loss term can be written as
 $(\md e/\md t)_{\rm c} =$ $-S$ where $S$ is given by ($p\to p/m_{\rm p}$): 
\eqb
S=-\frac{2}{3}\pi\, m_{\rm p} c^2 \,\frac{\partial u}{\partial x} 
     \int\limits_{0}^{\infty}
     \frac{\partial \tauesc(p)}{\partial p} p^5 f(p) \,\md p\,.
\eqe
Here we have used $\tauesc(0)=0$, $\tauesc(\infty)=1$, 
and $\partial\tauesc/\partial p\equiv 0$
for momentum $p\gg 1$, which is true, of 
course, for the representation given in Eq.~(\ref{tau}). Given a step function
$\tauesc(p)=H(p-p_0)$ the injection energy loss term used by (e.g.) 
Kang \& Jones
(1995) is revealed. The escape function $\tauesc$ depends on
the downstream plasma velocity, which is averaged over the diffusion length 
of the 
particles with momentum at the injection threshold. This dependence is quite
important for the injection efficiency, and leads to a regulation mechanism
similar to the above beam wave interaction. If the initial injection is
so strong that a significant amount of energy is transferred from the gas
to high energy particles, the downstream plasma cools, and becomes decelerated.
Because the injection momentum is in the high energy cut-off of the Maxwell
distribution of the plasma, the cooling decreases significantly the injection
rate. However, the deceleration allows for a modest increase of the phase space
of particles which can be injected. This is expressed by the $u_2$ dependence 
of Eq.~(\ref{tau}), and can be seen 
also in Fig.~\ref{tau_plot} where the dotted line shows the 
escape function for the setup distribution, and the solid line at $t=10\,t_0$. 
This velocity dependence balances partly the
reduction of injection due to the cooling of the plasma. These two effects 
lead to a very {\em weak} dependence of the injection efficiency on $\ep$ 
in the vicinity of $\ep\approx 0.35$.

We consider here diffusion proportional to Bohm diffusion, with 
 $\kappa=10\cdot\kappaB$, where
 $\kappaB=3\cdot10^{22}p^2/(1+p^2)^{1/2}\, {\rm cm}^2\,\rm{s}^{-1}$,
 for a magnetic field $B = 1\,\mu$G. 
Figure \ref{plasma} shows the gas density $\rho$, gas pressure $P_{\rm g}$,
plasma velocity $u$ and the cosmic ray pressure $P_{\rm c}$ over the spatial
length $x$, for different times. 
The scales are as follows: $t_0=1.2\cdot 10^5$ s,
$x_0=6.0\cdot 10^{13}$ cm, $u_0=5000\,{\rm km\,s}^{-1}$,
 $\rho_0/m_{\rm p} = 1\,{\rm cm}^{-3}$, 
$P_{\rm g 0}=4.175\cdot 10^{-7}\,{\rm erg\,cm}^{-3}$. The initial cosmic ray
adiabatic index is equal to the gas adiabatic index
 $\gamma_{\rm c} = \gamma_{\rm g} = 5/3$, and the compression ratio
is $r=3.97$. 
We have used 20480 grid zones for $x/x_0=[-18,14]$, with the shock initially 
at $x=0$, and 128 grid zones in $\log(p/m)=[-3.0,0.3]$.

\begin{wrapfigure}[22]{r}{8.5cm}
\vspace{7.8cm}
\includegraphics{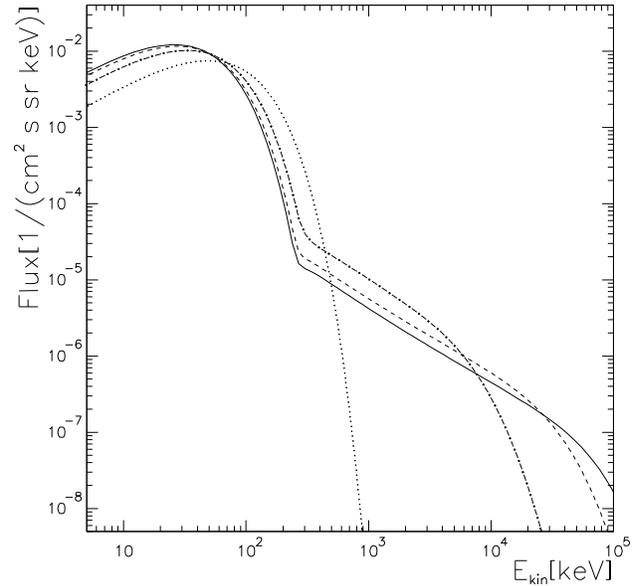}
\caption{\it Omni-directional flux vs.~proton kinetic
energy, for $t=0$ (dotted), $t=t_0$ (dot-dashed), $t=5\,t_0$ (dashed), 
and $t=10\,t_0$ 
(solid line)}\label{flux}
\end{wrapfigure}

For the parameters introduced above, Fig.~\ref{flux} shows the energy 
spectrum in form of the (at $t=0$) normalised omni-directional flux
 $F(E)\md E \propto v\,p^2 f(p)\md p$ vs.~proton kinetic 
energy downstream of the shock. At energies
above the thermal particles we expect, for the strong shock ($r\simeq 4$)
simulated here, the result $F(E)\propto E^{-\sigma}$, with
 $\sigma=\{(r+2)/(r-1)\}/2=1$, which 
is reproduced with high accuracy. At the thermal part of the distribution
the cooling of the plasma due to the energy flux into high energy particles
is responsible for the shift of the distribution towards lower energies.
The initial injection rate decreases thereby to a quite stable 
value, as described above.
Due to the steep dependence of both the Maxwell distribution, 
and the 
transparency function (Fig.~\ref{tau_plot}) on particle energy, the injection
energy is quite 
well defined, leading to a power law
due to Fermi acceleration, starting shortly above thermal energies.
In using the standard cosmic ray transport equation, we have, of course, made
use of the diffusion approximation, which may introduce an error especially for
$v\simeq u_2$.
Multiplication of the initial thermal distribution with 
$\tauesc$ suggest
an effective initial injection velocity of 
about $7\cdot 10^6\,{\rm m\, s}^{-1}$ (in the shock frame). 
Using an eigenfunction method, Kirk \& Schneider (1989) have explicitly
calculated the angular distribution of accelerated particles and accounted for
effects of a strong anisotropy especially at low particle velocities.
They were able to calculate the injection efficiency without recourse to
the diffusion approximation, and found always lower efficiencies.
For the above given injection velocity, $r=4$ and 
$u_0=5\cdot 10^6\,{\rm m\, s}^{-1}$, they estimate
a reduction effect of $\approx 7\%$, leaving the diffusion approximation 
as quite reasonable even in this regime.
\section{Conclusions}
We have presented results from a solution of the time dependent gas dynamics 
equation together with the cosmic ray transport equation. We have incorporated
in these calculations an analytical solution of an injection model for a
quasi-parallel shock, based on particle interaction with self-generated waves. 
We were therefore able to 
investigate the interaction of high energy particles, accelerated by the 
Fermi process, with the shocked plasma flow {\em without} a free parameter 
for the efficiency of the injection. 
We found the energy-flux $E\cdot F(E)$ of (non-relativistic) particles 
in the power law region to be about two orders of magnitude less than at the 
peak of the thermal distribution.
This result turns out to be quite stable, due to the self-regulating mechanisms
between particle injection and wave generation {\em and} gas modification
described above.

\section{Acknowledgments}
We are grateful to M.A. Malkov for helpful discussions. This work was 
supported by the University of Minnesota Supercomputing Institute, by NSF grant
AST-9619438 and by NASA grant NAG5-5055.
%
\vspace{1ex}
\begin{center}
{\Large\bf References}
\end{center}
Harten A., 1983, J. Comput. Phys. 49, 357\\
Kang H., Jones T.W., 1995, ApJ 447, 944\\ 
Kirk J.G., Schneider P., 1989, A\&A 225, 559\\
Malkov M.A., 1998, Phys. Rev. E 58, 4911\\
Malkov M.A., V\"olk H.J., 1995, A\&A 300, 605\\
Malkov M.A., V\"olk H.J., 1998, Adv. Space Res. 21, 551\\
Skilling J., 1975, MNRAS 172, 557\\
\end{document}